\documentclass[
    ,final            
  ,numberedheadings 
  ]
  {aipproc}
\usepackage{psfig}
\usepackage{graphicx}

\layoutstyle{6x9}

\begin{document}

\title{Search for narrow energy-shifted lines in XMM-Newton AGN spectra}


\author{A.L.Longinotti, S.Bianchi, M.Guainazzi, J.Roa-Llamazares, M.Santos-Lleo}{
  address={European Space Astronomy Centre, Apartado 50727 E-28080, Madrid, Spain}
}

\begin{abstract}
The detection of  X-ray narrow spectral 
features in the 5-7 keV band is becoming increasingly more common  in  AGN observations,
 thanks to the capabilities of current X-ray satellites.
Such lines, both in emission and in absorption, are mostly interpreted as 
arising from Iron atoms. 
When observed with some displacement from their rest frame position, 
these lines carry the potential to study the motion of circumnuclear gas in AGN, providing a 
diagnostic of the effects of the 
gravitational field of the central black hole.
These narrow features have been often found with marginal statistical 
significance.
We are carrying on a systematic search for narrow features using   spectra of bright type1 AGNs
 available in the {\it XMM-Newton} archive.
The aim of this work is  to characterise the occurrence of the narrow features phenomenon on a large sample of objects and to   estimate the significance of the features through Monte Carlo simulations.
The project and preliminary results are  presented.
\end{abstract}

\maketitle


\section{Introduction}
The presence of Fe~K features in the spectra of AGN and the interpretation 
on their origin are nowadays quite  well established.  
Fe K$\alpha$ transitions give rise to  fluorescence emission lines. The neutral Fe~K line at 6.4~keV is usually 
the most prominent in AGN spectra. An unresolved component of this transition is almost ubiquitous in  XMM-Newton
 AGN spectra (Bianchi et al.~2004, Page et al. 2004, Jim{\'e}nez-Bail{\'o}n et al. 2005)
 and it  is generally interpreted as being emitted via X-ray reflection on a distant medium, like the AGN torus 
(Ghisellini et al. 1994).
In some cases it is also possible to observe Fe~K  lines from ionised Iron atoms emitted  at 6.7 and 6.97~keV
(see left panel of Fig.1 for an example of Fe K lines complex in the Sy 1 Mrk~590). The origin of these lines is ascribed to warm gas surrounding the accretion disc.  
When Fe~K emission is produced  in the innermost region of the AGN, relativistic effects 
intervene and they can strongly modify the line profile, giving rise to a 
broad and skewed  Fe~K line  (Fabian et al.~2000).
This was the general picture up to few years ago. 
In fact the debate on the  nature of the inner accretion flow in AGN has been reinforced 
by the recent discoveries of narrow energy-shifted spectral features in the spectra of several Seyfert~1 Galaxies. 

 Narrow energy-shifted {\it emission lines} have been 
detected in the hard X-ray spectra of some AGN: 
NGC~3516 (Turner et al. 2002, 
Iwasawa et al.~2004), Mrk~766 (Turner et al.~2004, 2006),  ESO~198-G24 (Guainazzi 2003),
 NGC~7314 (Yaqoob et al.~2003), Mrk~841 (Petrucci et al.~2002, Longinotti et al.~2004),   ESO~113-G010 (Porquet et al.~2004),  
 UGC~3973, (Gallo et al.~2005).
An example of a redshifted emission line detected in NGC~3516 is shown in the right panel of Fig. 1. 
Theoretical models have predicted the possibility that Fe~K emission lines 
from the disc can be observed with a narrow profile under some conditions.
In fact, the broad  Fe line profile arises  by integrating the emission from a large area of the disc.
But, if the X-ray reflection arises as a result of magnetic flares in  localized regions on the disc, energy-shifted narrow lines are instead expected to be observed  in the spectrum. 
Nayakshin et al.~(2001) proposed a model in which the accretion disc is 
illuminated by magnetic flares inducing active regions (hot spots)
with limited size.
As these regions orbit around the black hole, the emitted line should appear 
in a range of energies  between $\sim$~4--8~keV   due to the Doppler and gravitational  effects.
If the emitting region co-rotates with the disc, the flux and the centroid energy of the line  
vary as a function of the orbital phase. As demonstrated by Dovciak et al.~(2004), the line profile computed in time-resolved intervals does show such variations 
and in theory, it is possible to reconstruct the path of the hotspot on the disc.

Another scenario in which narrow energy-shifted lines can be produced has been proposed by Turner et al.~(2002, 2004).
Redshifted lines are interpreted as being emitted by material which is either
being ejected at relativistic velocity on the far side or  falling towards the black hole.
The different ranges of energies of the line peaks would then reflect 
respectively a deceleration or acceleration  in the velocity of the gas due to the gravitational force. 
Remarkably, the ``aborted jets'' model recently proposed by Ghisellini et al. (2004) to explain the X-ray emission in radio-quiet AGN, predicts the presence of  blobs of gas in fast motion close to the black hole.

\begin{figure}
\psfig{figure=longinotti_1_fig1.ps,height=5.0cm,width=7cm,angle=-90}
\psfig{figure=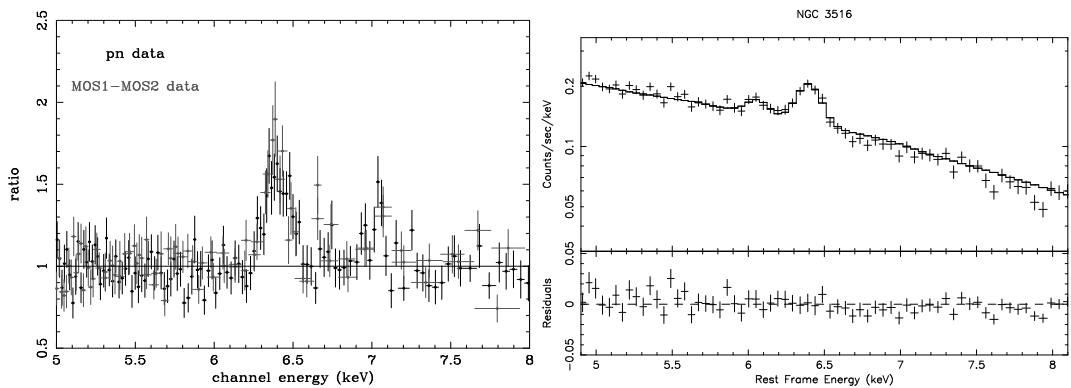,height=5.0cm,width=7cm,angle=-90}
\caption{Left side: The Fe K line complex in Mrk~590 as observed by  XMM-Newton (Longinotti et al. 2006a): three emission lines were detected at energy fully consistent with emission from FeI, FeXXV and FeXXVI (rest energies of 6.4, 6.7 and 6.97~keV). Right side: Fe K line in NGC 3516. On the red side of the 6.4~keV emission line, an emission feature was detected at $\sim$~6.1~keV (see Bianchi et al. 2004 for more details).}
\label{fig1}
\end{figure}

X-ray  {\it absorption lines} in the Fe K band may also provide insights on the central region of AGN.
The first detection of an absorption  redshifted Fe~K line  
was found in {\it ASCA} data of NGC3516 (Nandra et al.~1999). 
With the advent of {\it XMM$-$Newton} and {\it Chandra}, the number of absorption  
features in active galaxies spectra have considerably increased:
Q0056-363 (Matt et al. 2005),  E~$1821+643$ (Yaqoob \& Serlemitsos 2005, but see also Jim{\'e}nez-Bail{\'o}n et al. 2006), Mrk 509 (Dadina et al. 2005), PG$1211+143$ (Reeves et al. 2005), Mrk 335  (Longinotti et al. 2006b).
One of the best example of variable absorption structures from K$\alpha$ and K$\beta$ transitions of ionised Iron is the XMM-Newton spectrum of NGC~1365 (Risaliti et al. 2005) displayed in the left panel Fig. 2, whereas on the right side of the same figure
is plotted the redshifted absorption line detected in Mrk~335 (see caption for details).
   Several  cases of X-ray blueshifted absorption lines have been interpreted as evidence for  high velocity outflows: APM~08279+5255 and  PG1115+080 (Chartas et al.~2002, 2003),  PG1211+143 (Pounds et al.~2003), PDS~456 (Reeves et al.~2003), although  few of them somewhat questionable (McKernan et al.~2004). 
 The outflows interpretation is also  supported  by theoretical models where synthetic spectra  produced through radiative transfer calculations are found in
  good agreement with the observations on a qualitative level (Sim 2005).
According to accretion models,  high velocity gas should be present in the surroundings 
of the central region of AGN and the presence of such gas is in principle observable in the form of absorption features imprinted on the X-ray spectra (Crenshaw et al.~2003).
Depending on the kinematic structure and dynamical state of the gas, the absorption lines would be observed with different widths and with some displacement from their rest frame energy.
This gas is likely to be highly ionised due to the nuclear continuum irradiation and therefore is bound to produce absorption mainly in the Fe~K shell band.  Resonance transitions from FeXXV and FeXXVI produce absorption features at the energy of 6.7 and 6.97~keV. Taking into account the source redshift, these features should be observed in the $\sim$5--8~keV range for most objects. 
To be observed with considerable energy-shift, the lines must be forming in a plasma 
 infalling on the accretion disc or or outflowing at high velocity.
Although the exact nature of the energy shift of such lines is as yet unclear, the most 
likely scenario for producing the observed features  would involve a combination of gas 
orbiting in highly relativistic motion  and/or gravitational shifts of the photons (Ruszkowski et al.~2000).

\begin{figure}
\psfig{figure=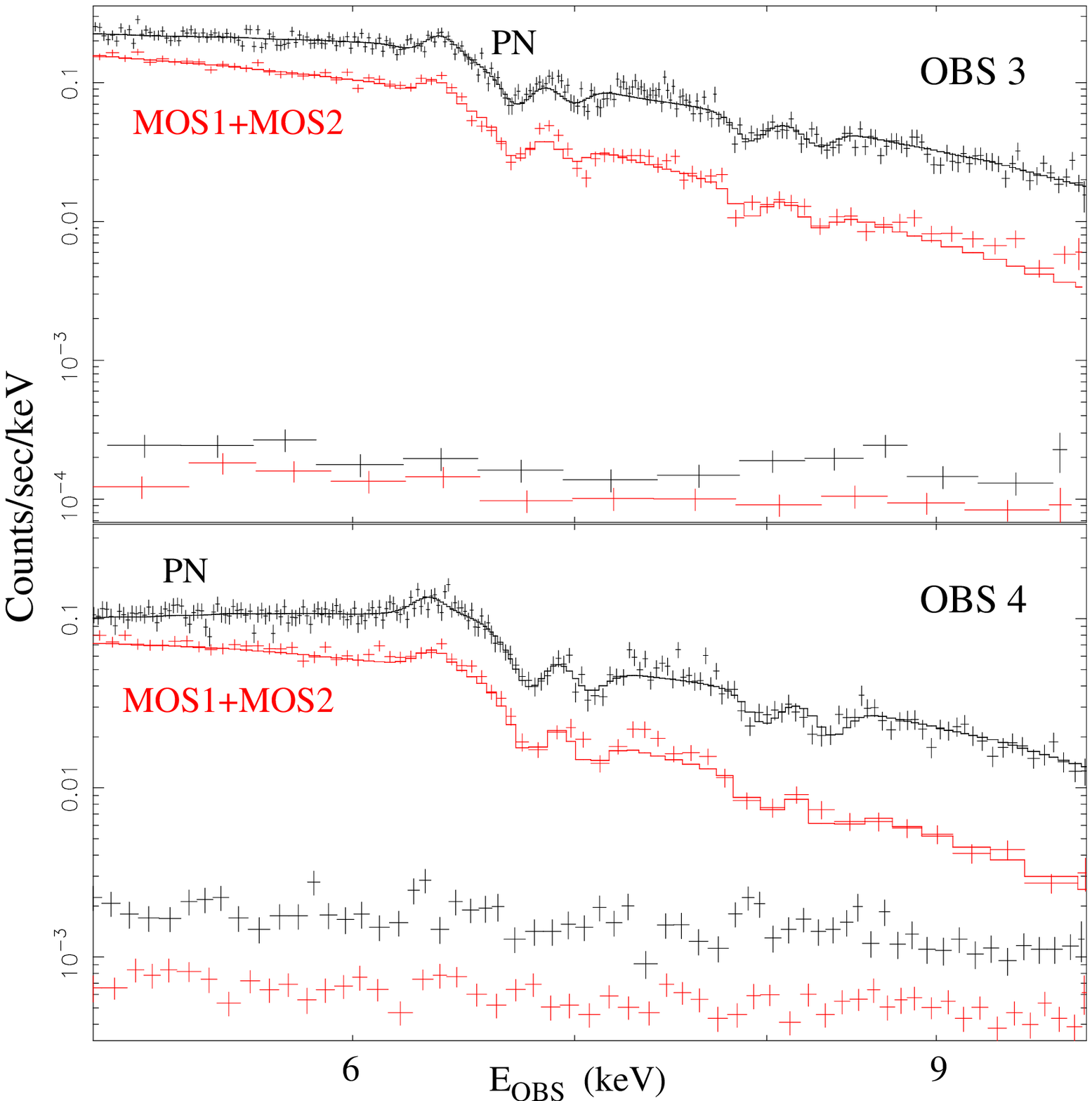,height=5.0cm,width=7cm}
\psfig{figure=longinotti_1_fig4.ps,height=5.0cm,width=7cm,angle=-90}
\caption{Left side: EPIC spectra of 2 distinct observations of NGC~1365, where variable absorption features from Fe XXV-XXVI have been detected at $\sim$ 7 and 8 keV (Risaliti et al. 2005). Right side:  The plot shows the residuals (data to model ratio) of the EPIC spectra for the Seyfert 1 galaxy Mrk~335, where a redshifted absorption line has been detected. The data are fitted by a power law with $\Gamma$= 2.15$^{+0.04}_{-0.03}$ and the  narrow absorption  line visible  at $\sim$5.9~keV is detected in this source with more than 99\% of significance. If 
interpreted as Fe XXVI resonance absorption, the shift implies a velocity of $\sim$0.11-0.15 {\it c} for the inflowing gas  (Longinotti et al., 2006b).}
\label{fig2}
\end{figure}

\section{Systematic search for narrow features}
 Emission and absorption energy$-$shifted lines are clearly two distinct phenomena, but
 they share some common properties:
 \begin{itemize}
 \item They are detected with marginal significance (2$-$3~$\sigma$)
 \item They have been found in individual objects, by individual authors and therefore 
 with various methods of data reduction and analysis, and with heterogeneous calibrations
  \item They are probably of transient nature, since most detections 
 have been  found in time$-$resolved spectra.
 \item  High confidence detections  of such  features would be of crucial importance 
 in testing  the black hole paradigm for AGN and would provide a new additional 
 tool to be used alongside the broad Fe~K$\alpha$ line.  
 \end{itemize} 
The project described in the following aims to perform a systematic search on narrow energy-shifted lines in the  {\it XMM-Newton} archive and to characterise the significance of the detections with Monte Carlo  simulations.

The data have been collected from the {\it XMM-Newton} archive.
 The spectra are part of a larger sample of AGN including all the observations  flagged as ``AGN'' in  the  {\it XMM-Newton} archive and public up to March 2006 (Bianchi et al. in preparation).
 We have selected only the sources with number of counts in the hard X-ray band 
 higher than 1000 counts, obtaining a list 
 of 124 spectra extracted from  85 type-1 AGNs (multiple observations are available for many sources). 
Table 1 reports the optical classification of the objects. 
 The data have been processed with SAS version 6.5.0 and analysed with XSPEC 12.2.1ao.


\begin{table}
\begin{tabular}{cc}
\hline
Object & Number \\
type   &         \\  
\hline
Seyfert 1 & 31  \\
Radio Quiet QSO & 39\\
Narrow Line Sy1 & 15\\
\hline
\end{tabular}
\caption{Object type in our sample}
\label{tab:a}
\end{table}

\subsection{Methodology}
The procedure described hereafter  is adapted from Longinotti et al. (2006b), where it was used to test the significance  of the redshifted absorption line in Mrk~335.
For the present work, each spectrum in the sample has been fitted in the 2-10~keV range with a baseline model defined by a power law
+ 3 Gaussian emission lines with energies at 6.4, 6.7, 6.97~keV and width fixed to 1~eV.
Then, the presence of additional narrow lines is tested by adding 
 {\it n} narrow Gaussian lines to the baseline model, allowing positive and negative 
 deviations for the line intensity and excluding the range 6.4-7~keV.
 The energy of these lines is free to vary between 4 and 9~keV.
 In this way a best fit model made by a power law+3~Fe~K lines +  {\it n}~Gaussian lines
 is found.
The significance of these additional Gaussian lines with respect to the 
best fit model  is checked through  the F-test.
Only those lines that are found  significant at more than 99\% for the F-test are included in the best fit (this conservative threshold has been arbitrarily chosen).  

In the next step we test the significance of the {\it n} Gaussian lines in the best fit.
 A  line detected in this way could be due to  random deviations originated by Poisson noise in the data.
Thus, the question is  `` what is the probability that a deviation in $\chi^2$ this large or larger
is  obtained by chance ?''.
 To answer it, we run Monte Carlo simulations on each spectrum where a shifted line has been detected at more than 99\% for the F-test.
The algorithm for the simulations is as follows.
We produce 10$^4$ realisations of the spectrum fitted with an input model 
made by the best fit model without the line that is being tested. 
In this way we obtain 10000 fake background-subtracted  data sets with photon statistics and spectral shape corresponding 
to the original spectrum.
Each of the simulated spectra is fitted with the input baseline model. 
Then, the Gaussian line to be tested is added back to the baseline model, a new spectral fit is performed  and the improvement 
in  $\chi^2$ for adding that line is recorded for each of the simulated spectra.
The number of $\Delta\chi^2$ larger than the one found in the real spectrum  will be used to estimate  the significance of the line in the real spectrum. 

\subsection{Preliminary results and considerations on the method}
At the time of the writing, only few test-runs have been performed, setting the threshold of significance at 99 and 95\%, in order to understand how the method works.
This  allow us to identify some drawbacks of the method and improve it  before   reporting unclear or incomplete results.
 Some of these problems are highlighted in the following:
 the method does not take into account the complexity of many AGN spectra, particularly in Seyfert Galaxies. For example, if a broad line or spectral curvature is present, 
 the automatic procedure would mimick the broad feature with many adjacent narrow lines.
 The transient nature of the shifted features constitutes another issue to be taken into account and solved: in a systematic search on the integrated spectra, some of the transient feature would be lost, as it happens in our sample for NGC~3516.
 However, a preliminary check on the well known objects PG1211+143 and Mrk~509 where detections of narrow lines have been claimed, confirms the 
presence of shifted lines and therefore we are confident that our procedure provides a good basis to search for shifted lines systematically. 

Among the many  future developments of  this work, 
we foresee to  test the detection procedure varying the baseline models 
for the simulated  spectra.
We would like  to run it   in samples of different objects
in the {\it XMM-Newton} archive (e.g. Blazars) to be used as a control sample
 and to take into account 
the transient nature of the features performing the search in time-resolved spectra.




\end{document}